# Structural, magnetic and superconducting phase transitions in $CaFe_2As_2$ under ambient and applied pressure.

P. C. Canfield,[1] S. L. Bud'ko,[1] N. Ni,[1] A. Kreyssig,[1] A. I. Goldman,[1] R. J. McQueeney,[1] M. S. Torikachvili,[2] D. N. Argyriou,[3] G. Luke,[4] and W. Yu[4]

1) Ames Laboratory and Department of Physics and Astronomy, Iowa State University, Ames, Iowa 50011, USA

2) Department of Physics, San Diego State University, San Diego, California 92182-1233, USA

3) Helmholtz-Zentrum Berlin fűr Materialien und Energie, Glienicker Str. 100, 14109 Berlin, Germany

4) Department of Physics and Astronomy, McMaster University, Hamilton, Ontario L8S 4M1, Canada

## Abstract

At ambient pressure $CaFe_2As_2$ has been found to undergo a first order phase transition from a high temperature, tetragonal phase to a low temperature orthorhombic / antiferromagnetic phase upon cooling through T ~ 170 K. With the application of pressure this phase transition is rapidly suppressed and by ~ 0.35 GPa it is replaced by a first order phase transition to a low temperature collapsed tetragonal, non-magnetic phase. Further application of pressure leads to an increase of the tetragonal to collapsed tetragonal phase transition temperature, with it crossing room temperature by ~ 1.7 GPa. Given the exceptionally large and anisotropic change in unit cell dimensions associated with the collapsed tetragonal phase, the state of the pressure medium (liquid or solid) at the transition temperature has profound effects on the low temperature state of the sample. For He-gas cells the pressure is as close to hydrostatic as possible and the transitions are sharp and the sample appears to be single phase at low temperatures. For liquid media cells at temperatures below media freezing, the $CaFe_2As_2$ transforms when it is encased by a frozen media and enters into a low temperature multi-crystallographic-phase state, leading to what appears to be a strain stabilized superconducting state at low temperatures.

## Introduction

The discovery of superconductivity in RFeAsO [1] (R = rare earth) and $AEFe_2As_2$ (AE = Ba, Sr, Ca) materials [2] is further evidence that the hunt for new superconductors is alive and well, and still is, fundamentally, a highly intuitive, experimental endeavor. Whereas the RFeAsO materials manifest higher $T_c$ values [3], approaching 60 K, they have proven difficult to grow as single crystals of and suffer from a plethora of synthetic problems associated with control and evaluation of oxygen and fluorine (when used) stoichiometry. The $AEFe_2As_2$ materials, on the other hand, were quickly grown in single crystalline form [4] first out of Sn and later out of FeAs.[5, 6] Although their transition temperatures are closer to that of $MgB_2$ [7, 8] than to the highest values found in the F-doped RFeAsO compounds, the salient physics appears to be the same, or at least similar, and more readily available single crystals, as well as potentially better and more quantifiable control over substitution, has made the $AEFe_2As_2$ materials subject to extensive study.

Reports of superconductivity in K-doped $BaFe_2As_2$ [2] and $SrFe_2As_2$ [9] (both know examples of the $ThCr_2Si_2$ structural family [10]) was followed by the discovery of a new member of this structural class: $CaFe_2As_2$ [11] and the stabilization of superconductivity in it via Na-doping [12]. In many ways $CaFe_2As_2$ epitomizes the salient physical features of the $AEFe_2As_2$ compounds and serves as a readily accessible model system. In other ways it provides several important warnings and has alerted the community to the complexities that appear to be inherent to these fascinating compounds.

This article will attempt to review the remarkable progress that has been made in understanding the nature of the structural, magnetic and electronic states of $CaFe_2As_2$ under ambient and applied pressure. As such it will review recent work about growth and basic structural, thermodynamic and transport properties of $CaFe_2As_2$, [11-13] details of the structural and magnetic transitions at ambient pressure [14] as well as under applied pressure [15,16], and the effects of applied pressure on the transport and thermodynamic properties of $CaFe_2As_2$. [17-20]

**Synthesis and basic properties**

$CaFe_2As_2$ was first synthesized and identified in single crystalline form. [11] Given that single crystals of $BaFe_2As_2$ and $SrFe_2As_2$ were successfully synthesized from Sn flux, [4, 21] $CaFe_2As_2$ was a logical "next step" up the alkali-earth column of the periodic table. We have found that a starting stoichiometry of $Ca_{2.13}Fe_{3.74}As_{4.28}Sn_{89.84}$ produces the best, square-plate-faceted samples. The starting elements are placed in an $Al_2O_3$ crucible and sealed in amorphous silica tubes, as described in refs. 11 and 22, and cycled through the following schedule: 6 hours to heat from room temperature to 600 °C, a 1 hour dwell, then 4 hours to heat to 1150 °C, 4 hours dwell, then 50 hours cooling to 600 °C, followed by decanting. The small amount of remaining Sn flux on the resulting single crystals is removed by etching in concentrated HCl for 10 to 20 seconds. Clean single crystals, with planar dimensions as large as 5 x 5 $mm^2$, can be readily found. A representative sample is shown pictured over mm-grid graph paper in the inset to Fig. 1.

Temperature dependent electrical resistivity and magnetic susceptibility measurements [11] both show a clear and sharp phase transition at ~170 K, as shown in Figs. 1 and 2. It is worth noting that these features are remarkably robust and appear in virtually the same form and detail in all of the original reports on $CaFe_2As_2$. [11-13] The planar ($\rho_{ab}$) resistivity decreases with decreasing temperature from a room temperature value near 0.3 mΩ cm to near 0.05 mΩ cm at 2 K, with a discontinuous, and hysteretic jump in resistivity upon cooling (warming) through 170 K transition. The magnetic susceptibility decreases very weakly with decreasing temperature (in both directions of applied field) but manifests a dramatic decrease upon cooling through the transition temperature. Specific heat data (Fig. 3) show a sharp feature at this temperature, consistent with a latent heat that appears slightly broadened by the measurement technique. [11] The low temperature electronic specific heat can be evaluated from a plot of C/T vs. $T^2$ (Fig. 3, lower inset) and is ~ 5 mJ/mole $K^2$, consistent with a relatively low density of states at the Fermi-level in the low-temperature state.

**Structural determination at ambient pressure**

The thermodynamic and transport data shown in Figs. 1-3 are consistent with a first order phase transition near 170 K. To determine the precise nature of this transition, a single crystal x-ray diffraction study was performed. [11] ($\xi$ $\xi$ 0) and (0 0 $\xi$) scans through the (0 0 10) reflection (Fig. 4) show no change in the shape of the (0 0 10) reflection. However, in a narrow temperature range between 168.5 K and 171 K we observe coexistence of two reflections in longitudinal (0 0 $\xi$) scans. The reflection at lower ξ values in the (0 0 ξ) scans corresponds to the (0 0 10) reflection of the low temperature phase. The reflection at higher ξ values represents the high-temperature phase. In transverse (ξ ξ 0) scans, however, only one sharp reflection was observed. The abrupt change in the *c*-lattice parameter, together with the narrow temperature range of coexistence of both phases, indicates a first order phase transition at T ~ 170 K. This interpretation is corroborated by our measurements of the (1 1 10) reflection. As shown in Fig. 5, below 170 K, a splitting of the (1 1 10) reflection was observed in (ξ ξ 0) scans, consistent with a tetragonal-to-orthorhombic phase transition with a distortion along the diagonal (1 1 0) direction. Between 168.5 K and 171 K, reflections related to both phases coexist. The central peak corresponds to the (1 1 10) reflection of the tetragonal high-temperature phase. The surrounding pair of reflections arises from the two sets of 'twin domains' that arise from the orthorhombic distortion along the diagonal (1 1 0) direction for the low-temperature phase.

By analyzing the position of the (0 0 10) reflection in longitudinal (0 0 ξ) scans, the *c*-lattice parameter can be determined (Fig. 6). The in-plane *a*- and *b*-lattice parameters (Fig. 6) have been calculated based on the distance between the reflections close to the tetragonal (1 1 10) position and the (0 0 10) reflection in transverse scans along the (ξ ξ 0) direction. Between 10 K and 170 K, the difference in the orthorhombic *a*- and *b*-lattice parameters decreases slowly and monotonically with increasing temperature whereas the *c*-lattice parameter increases monotonically up to 170 K. At 170 K, the *a*-lattice parameter jumps abruptly as can be readily seen from the transverse (ξ ξ 10) scans through the (1 1 10) reflection in Fig. 5 and the temperature dependence of lattice parameters in Fig. 6. The jump is as large as ~ 1% between the high-temperature tetragonal value and the averaged low-temperature in-plane lattice parameters *a* and *b*. This change is quite large and emphasizes the first order nature of the structural phase transition at T ~ 170 K. The *c*-lattice parameter shows also a jump at the transition temperature, but smaller than the in-plane lattice parameter.

The inset to Fig. 6 shows the intensity of the low- and high-temperature (0 0 10) reflection as a function of temperature close to the tetragonal to orthorhombic phase transition. The overall behavior reflects again the first order character of the phase transition as well as the high quality of the sample. Coexistence of the high-temperature tetragonal and low-temperature orthorhombic phase can be detected for a very narrow temperature range of only ~ 2 K, representing the 'sharpness' of the transition in the studied sample.

This high temperature tetragonal to low temperature orthorhombic, first order, structural phase transition is directly, and intimately, coupled with a first order transition to a low temperature, antiferromagnetically ordered state. This can be most clearly seen in the results of elastic neutron diffraction [14] shown in Figures 7 and 8. The antiferromagnetically ordered state (shown pictorially in Fig. 7) is a commensurate structure with a propagation wave vector of (0,1,0) associated with an ordered Fe moment of ~ 0.8 $\mu_B$ directed along the orthorhombic *a*- axis. The magnetic cell is the same as the chemical unit cell and the moments are ferromagnetically coupled along the *b*-direction and antiferromagnetically coupled along the **a**-and **c**-directions.[14]

The most interesting aspect of these elastic neutron scattering measurements deals with the behavior of the magnetic and structural orders close to the transition temperature, as illustrated in Fig. 8. Figure 8a plots the temperature dependences of both the orthorhombic splitting of the nuclear peak and the intensity of the magnetic peak (normalized to the volume fraction of the magnetic orthorhombic phase) upon warming and cooling through the transition. Here we see that the structural transition, as measured by the orthorhombic splitting, is discontinuous over a temperature range of less than 0.5 K. Figure 8b plots the observed intensities of the magnetic (−1 0 1) peak and the orthorhombic (400) nuclear peak on cooling through the transition region where there is coexistence between the low-temperature antiferromagnetic orthorhombic and high-temperature tetragonal phases. The volume fraction of the orthorhombic phase was calculated from fits to the composite (400)/(040) orthorhombic and (220) tetragonal peaks in this temperature range. Whereas the magnetic ordering appears to evolve continuously over this coexistence regime (Fig. 8b), normalizing these data by the orthorhombic phase volume fraction, which increases in the same manner with temperature, shows that the magnetic transition is, indeed, discontinuous at the structural transition (Fig. 8a). Figure 8a also displays a clear signature of hysteresis in the tetragonal-to-orthorhombic transformation over a range of approximately 1 K. Perhaps most interesting, however, is that the magnetic ordering shows this same hysteresis upon cooling and warming, clearly demonstrating that the structural transition and magnetic ordering are intimately connected.

**Properties of $CaFe_2As_2$ as a function of applied pressure**

Although the ambient pressure properties of $CaFe_2As_2$ appear to epitomize the general behavior of the undoped $AEFe_2As_2$ (AE = Ba, Sr, Ca) materials, [2,9] none of these parent compounds, at ambient pressure, allow for the study of superconductivity. Given that pressure (both physical and chemical) was key to raising the $T_c$ value of the F-doped RFeAsO materials [3, 23] it was reasonable to hope that $CaFe_2As_2$ should be the most promising of the $AEFe_2As_2$ materials for pressure stabilized superconductivity. Based on this premise, measurements were made that demonstrated that the application of hydrostatic pressures, at room temperature, via a liquid medium, self-clamping pressure cell of up to ~ 2.0 GPa lead to the suppression of the 170 K structural / antiferromagnetic phase transition and the stabilization of superconductivity at low temperature. [17]

Figure 9 presents the electrical transport data taken on single crystalline $CaFe_2As_2$ for applied pressures below 2.0 GPa.[17] There are several features worth noting in these data. First, the resistive signature of the ambient pressure transition to the low temperature, orthorhombic / antiferromagnetic state is suppressed, and above P ~ 0.35 GPa it vanishes. Secondly, for higher pressures a new feature, a broad loss of resistance upon cooling, appears and increases in temperature as pressure is increased, passing through room temperature for applied pressures of ~ 1.7 GPa. Third, at intermediate pressures, superconductivity is detected (Fig. 9b) and a dome of superconductivity can be inferred to exist between ~ 0.3 and 0.9 GPa. Fourth, the normal state, residual resistivity (approximated by the resistivity value measured at 15 K) manifests a dramatic pressure dependence, dropping precipitously near 0.5 GPa and saturating to a fixed, high pressure value for P > 0.8 GPa.

All of these observations are clearly illustrated in Fig. 10, the T-P phase diagram for the $CaFe_2As_2$ system (for pressure applied with a liquid medium cell). At higher temperatures, the ambient pressure phase transition is suppressed and essentially disappears between 0.3 – 0.5 GPa. At higher pressures a new phase (identified as the collapsed tetragonal phase below [15, 16]) emerges and has its transition temperature rise with further application of pressure. Precisely in the pressure region where these two transitions appear to intersect, a dome of superconductivity exists at lower temperatures. This dome is essentially centered on the pressure at which the low temperature resistivity drops precipitously. The key pressure region, below 1.0 GPa, is accessible to a wide range of measurements as well as pressure apparati. These results appear to be quite robust, with superconductivity being detected [18] and the resistively determined phase diagram being reproduced [19] (down to fine details) by other groups.

These data clearly raised a series of questions, primarily associated with the nature of the high pressure, low temperature state as well as the nature of the phase transitions into it, either as a function of temperature or pressure. To address these questions a series of scattering studies as a function of pressure and temperature were made. [15, 16] Given the relatively low pressures needed for these studies, a He-gas pressure cell was used since it would allow for isothermal sweeps of pressure for T > 50

K. Figure 11 presents a summary of these data (as well as a higher pressure data point determining the pressure at which the high pressure phase transition crosses room temperature—see inset) in the form of a T – P phase diagram.[16] At a gross level, the high temperature, tetragonal phase was found to exist below room temperature for pressures up to ~ 1.7 GPa and down to temperature as low as ~ 120 K, for P ~ 0.3 GPa. At low pressures ( P < ~ 0.35 GPa) the low temperature phase is that described above for ambient pressure, i.e. orthorhombic and antiferromagnetically ordered, and at higher applied pressure (P >~ 0.35 GPa) a collapsed tetragonal phase was discovered.[15] This collapsed tetragonal phase does not manifest any detectable magnetic order and is thought to be non-moment bearing.[15,16] Whereas details about the crystallography as well as the magnetic structure can be found in refs. 15 and 16, it is worth noting that the collapsed tetragonal phase is associated with a dramatic change in the unit cell parameters of $CaFe_2As_2$. When the sample is cooled across the tetragonal - collapsed tetragonal phase transition there is an ~ 5% decrease in volume associated with an extremely anisotropic change in the unit cell dimensions: the *a*-axis expands by ~ 2.5 % and the *c*-axis contracts by ~ 9%. The hysteresis found at ambient and applied pressures [11-13, 17, 19] was detected in the scattering data and is shown as the grey shaded area in the T- P diagram (Fig. 11).

There is broad agreement between the high temperature phase lines shown in Fig. 10 and Fig. 11. In addition, a near vertical, first order phase line was detected in the He-gas cell data (as a result of varying pressure at fixed temperatures), separating the low temperature orthorhombic / antiferromagnetic phase from the higher pressure, collapsed tetragonal. Despite these similarities, there are also some disturbing differences between the transport and scattering data sets, the most conspicuous of which is the sharpness of the transitions, specifically the transition from the high temperature tetragonal phase to the low temperature, collapsed tetragonal phase. Figure 12 shows the (004) structural peaks for the tetragonal and collapsed tetragonal phases as a function of temperature for an applied pressure of 0.47 GPa (He-gas cell). There is a clear thermal hysteresis, as shown in Fig. 11 and found in transport data, [17, 19] but the transition is still quite sharp. On the other hand, the resistive transition seen in Fig. 9 for P ~ 0.55 GPa (liquid media cell) is ~ 40 K wide. This broadening of the transport features is also apparent when comparing the sharp resistive anomaly associated with the transition to the orthorhombic / antiferromagnetic state at ambient pressure to the broadened feature for the same transition at 0.23 or 0.35 GPa (Fig. 9), obtained with a liquid medium cell.

Given the remarkable pressure sensitivity of $CaFe_2As_2$ it became apparent that one possible origin of this difference between the scattering and transport data was the difference in the precise manner of pressure transmission to the single crystalline samples: liquid media for the initial transport measurements and He-gas media for the scattering measurements. Although both of these media are fluids at room temperature (at least for P below ~ 2.0 GPa), the liquid media used in the clamp cells for transport measurements freezes well above the 120 K minima in the tetragonal phase near 0.35 GPa. This means that when the sample undergoes the first order phase transition, with the accompanying sudden change in volume, the pressure media is no longer fluid and the pressure response is no longer hydrostatic. In specific, when the sample enters the collapsed tetragonal phase the sample is encased in a solid when it suddenly experiences a dramatic increase in the *a*- and *b*- axes and a dramatic decrease in the *c*-axis. This can easily result in non-hydrostatic components to the effective pressure on the sample, e.g. uniaxial and shear stresses. Such poorly controlled pressure gradients and components can easily lead to a broadening of the phase transition.

In order to check this hypothesis and also have data taken under similar conditions as the neutron scattering data, a He-gas pressure cell was used to provide the pressure environment for electrical resistivity and magnetic susceptibility measurements.[20] Temperature dependent resistivity data at representative pressures from this He-gas pressure cell are shown in Fig. 13. These data clearly support the hypothesis that the difference between the first transport study in a liquid medium cell and the diffraction studies in He-gas pressure cells is associated with the pressure media. As clearly displayed in Fig. 13a, the low pressure ($P < 0.36$ GPa) phase transition from the high temperature tetragonal to the low temperature orthorhombic / antiferromagnetic phase remains extremely sharp as it is suppressed. In addition the higher pressure ($P > 0.35$ GPa) phase transition into the collapsed tetragonal phase manifests a similar sharpness in the He-gas pressure cell. Indeed the sharpness of the feature associated with this transition as well as the hysteresis associated with the transition to (and from) the collapsed tetragonal phase is remarkably similar in the scattering and transport data when both are taken in He-gas cells. This can be seen by comparing the $P = 0.47$ GPa scattering data shown in Fig. 12 with the $P = 0.42$ GPa transport data in Fig. 13. It is worth noting that for one pressure ($P = 0.352$ GPa) all three structural phases are present, each over a limited temperature range, below room temperature. The P = 0.352 GPa data help define just how close to vertical the phase transition between the low temperature orthorhombic and collapsed tetragonal phases is (see Fig. 14 below).

Figure 13b plots the low temperature resistivity for various pressures on a semi-log scale, similar to the data presented in Fig. 9b. These data show another very sharp feature: the sudden change in the low temperature residual resistivity (again approximated by the 15 K resistivity) as a function of applied pressure. This is a consequence of the sharp drop in resistivity that accompanies the transition to the collapsed tetragonal phase in the He-gas cell transport data.

Figure 14 is the T-P phase diagram assembled from He-gas cell electrical resistivity and magnetic susceptibility data.[20] The resistivity at 15 K is plotted on the right hand axis (as it was in Fig. 10 for the liquid medium pressure cell data) to show the dramatic drop in resisitivity associated with the sharp transition into the collapsed tetragonal phase.

Although the He-pressure cell data reconciles the scattering and transport data very well, there is a conspicuous difference between the transport data in liquid media and in the He-gas cell: for the latter there is no transition to a zero resistance state at low temperatures. At best, there is a slight down turn in resistivity below 10 K for pressures very close to P = 0.33 GPa. Whereas there was a clear superconducting dome in the T-P phase diagram determined from liquid media pressure cell work, [17,19] there is no such dome in T-P phase diagram from the He-gas cell. This difference is consistent with the earlier correlation between the existence of the superconducting dome and the pressure region associated with the rapid change in low temperature resistivity. Whereas for the liquid medium cells this region spanned several tenths of GPa, for the He-gas cell this "region" may well be very narrow, if it has any width at all. For the He-gas cell, the dome may well have become a delta function, if it exists at all.

A final set of data can shed further light on the differences between the He-gas and liquid media cell results: neutron scattering data taken in a liquid media cell. Figure 15 further demonstrates the effects of cooling $CaFe_2As_2$ through its structural phase transition in a media that has already solidified.[16] Although the sample, at this pressure, should cleanly transform into the collapsed tetragonal phase, the fact that the phase transition involves an expansion in two directions and a dramatic contraction in the third indeed appears to lead to a multi-crystallographic-phase sample at low temperatures. In addition, the transition measured in Fig. 15 is does not occur sharply at a single temperature, but instead is spread over tens of Kelvin. This is in stark contrast to the scattering data taken under pressures generated by He-gas cells and consistent with the broadening seen in the liquid media transport data (Fig. 9). This smearing of the transition and establishment of a multi-crystallographic-phase sample was one of the possibilities discussed in a recent μSR study of $CaFe_2As_2$ under pressure [24]; a possibility that has subsequently been refined into a near certainty. [20]

**Conclusions and Summary**

At ambient pressure $CaFe_2As_2$ manifests a dramatic and strongly coupled first order phase transition from a high temperature, tetragonal phase to a low temperature, orthorhombic, antiferromagnetic phase as it is cooled below ~ 170 K. This phase transition, as well as the low temperature structure of $CaFe_2As_2$, is extremely sensitive to the application of pressure. When the sample is cooled *through* the structural phase transition under *hydrostatic* pressure there is a sharp, well defined, transition to either the lower pressure orthorhombic phase, or the higher pressure collapsed tetragonal phase. On the other hand, if the sample is embedded in a solid media at the temperature of the structural phase transition, then the phase transition is broadened, and, in the case of the transition to the collapsed tetragonal phase, this broadening can extend over tens of Kelvin. The non-hydrostatic conditions brought on by the combination of the freezing of the media and a first order phase transition with dramatic, anisotropic, changes in unit cell volume lead to a multi-crystallographic-phase sample existing over a wide temperature range. To this end, $CaFe_2As_2$ is an exquisite example of the importance of hydrostaticity not only at ambient temperature, but at all temperatures, especially through structural phase transition temperatures.

$CaFe_2As_2$ provides important insight into the superconductivity in the $AEFe_2As_2$ family of compounds. It further illustrates the importance of suppressing the transition from the high-temperature tetragonal state to low enough temperature so as to allow for the stabilization of the superconducting state. It is apparent from the He-gas cell data that when $CaFe_2As_2$ is deep within either the orthorhombic / antiferromagnetic state, or the collapsed tetragonal state (in which no long range magnetic order has been detected) superconductivity is not supported. These data, taken together with doping data (such as, but not limited to, Refs. 2, 3, 6, 9) indicate that superconductivity only appears to be stabilized when the compound is either in the higher temperature, tetragonal state or close to it. The fact that superconductivity can be stabilized in at least part of the $CaFe_2As_2$ sample under non-hydrostatic conditions (i.e. liquid media cells) implies that the superconductivity is most likely associated with either some residual tetragonal phase or some intermediate phase at the interface between the orthorhombic and collapsed tetragonal phases. This result brings up the possibility of strain stabilized superconductivity being induced and used in extruded powder-in-tube wires as part of the extrusion process as well as perhaps explaining some of the reports of measurements of superconductivity at ambient or low pressures in $SrFe_2As_2$ as well as $BaFe_2As_2$. Indeed recently this has been experimentally confirmed in $SrFe_2As_2$. [25]

**Acknowledgements:**

The breadth and detail of work accomplished on $CaFe_2As_2$ was possible because of the highly integrated and collaborative Ames Laboratory research environment that epitomizes the best of research at a National Laboratory. Work at the Ames Laboratory was supported by the Department of Energy, Basic Energy Sciences under Contract No. DE-AC02-07CH11358. M. S. T. gratefully acknowledges support of the National Science Foundation under No. DMR-0306165 and No. DMR-0805335. Research at McMaster University was supported by NSERC and CIFAR.

**Figures**

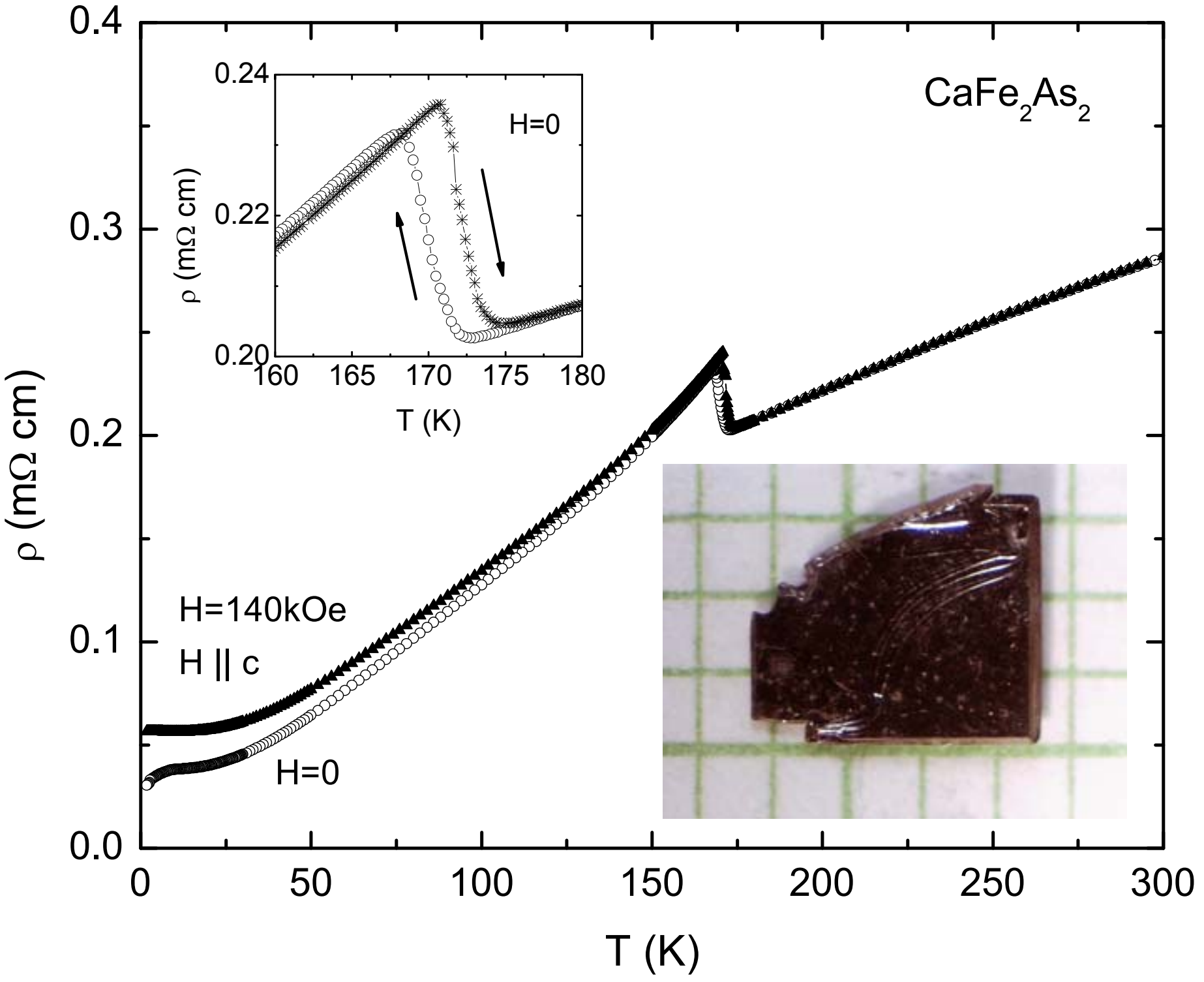

Figure 1: Temperature-dependent electrical resistivity of $CaFe_2As_2$, with current flowing within the basal plane, for *H* || *c*, *H*=0, and 140 kOe. Upper inset: hysteresis in temperature dependent resistivity near the 170 K phase transition (the arrows indicate increasing and decreasing temperature scans). Lower inset: picture of a $CaFe_2As_2$ single crystal on a millimeter grid paper. (After Ref. 11)

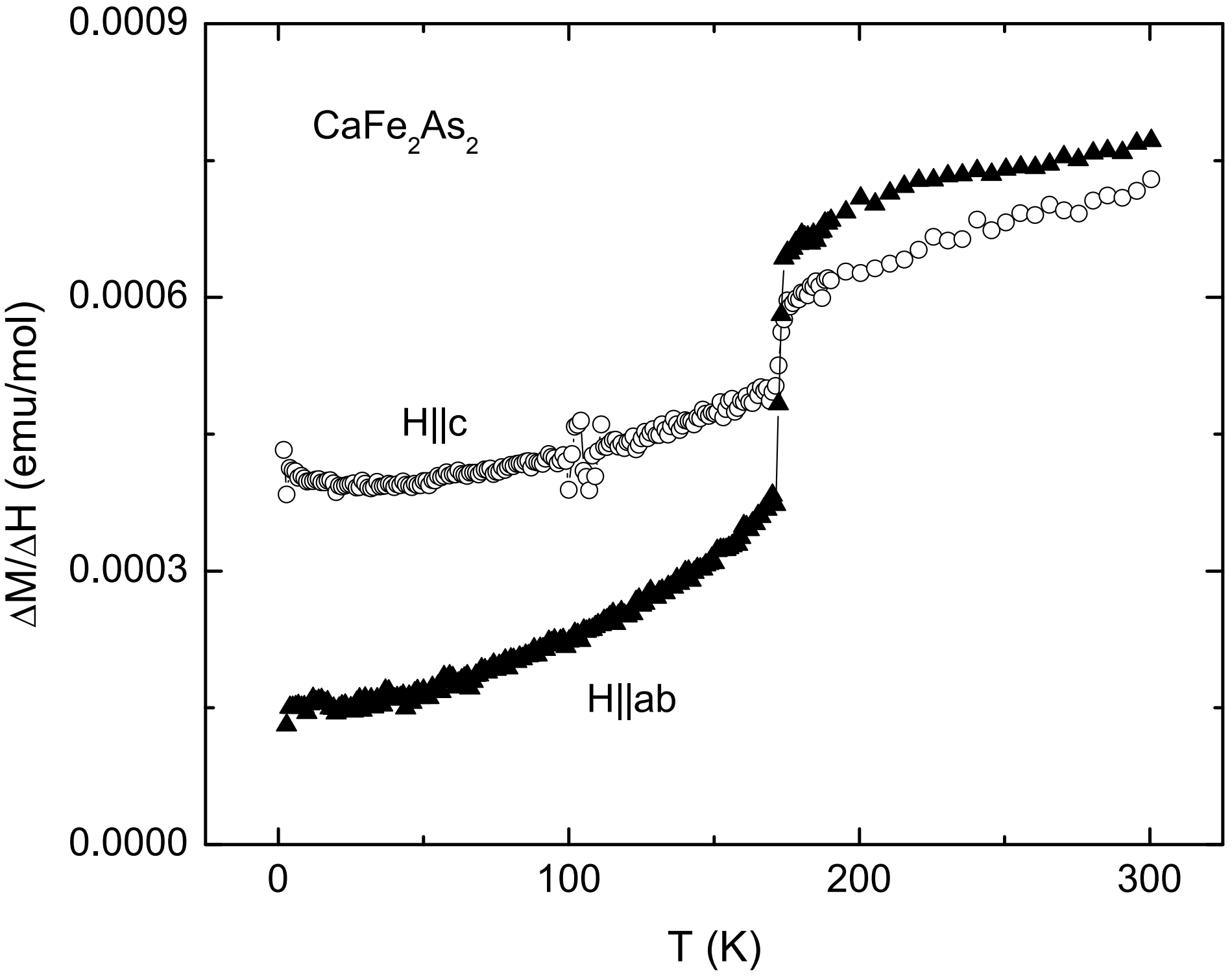

Figure 2: Temperature-dependent magnetic susceptibility of $CaFe_2As_2$ for applied field parallel to and perpendicular to the crystallographic *c* axis. The susceptibility was determined by the difference between 30 and 50 kOe magnetization runs so as to eliminate weak ferromagnetic contributions. (After Ref. 11)

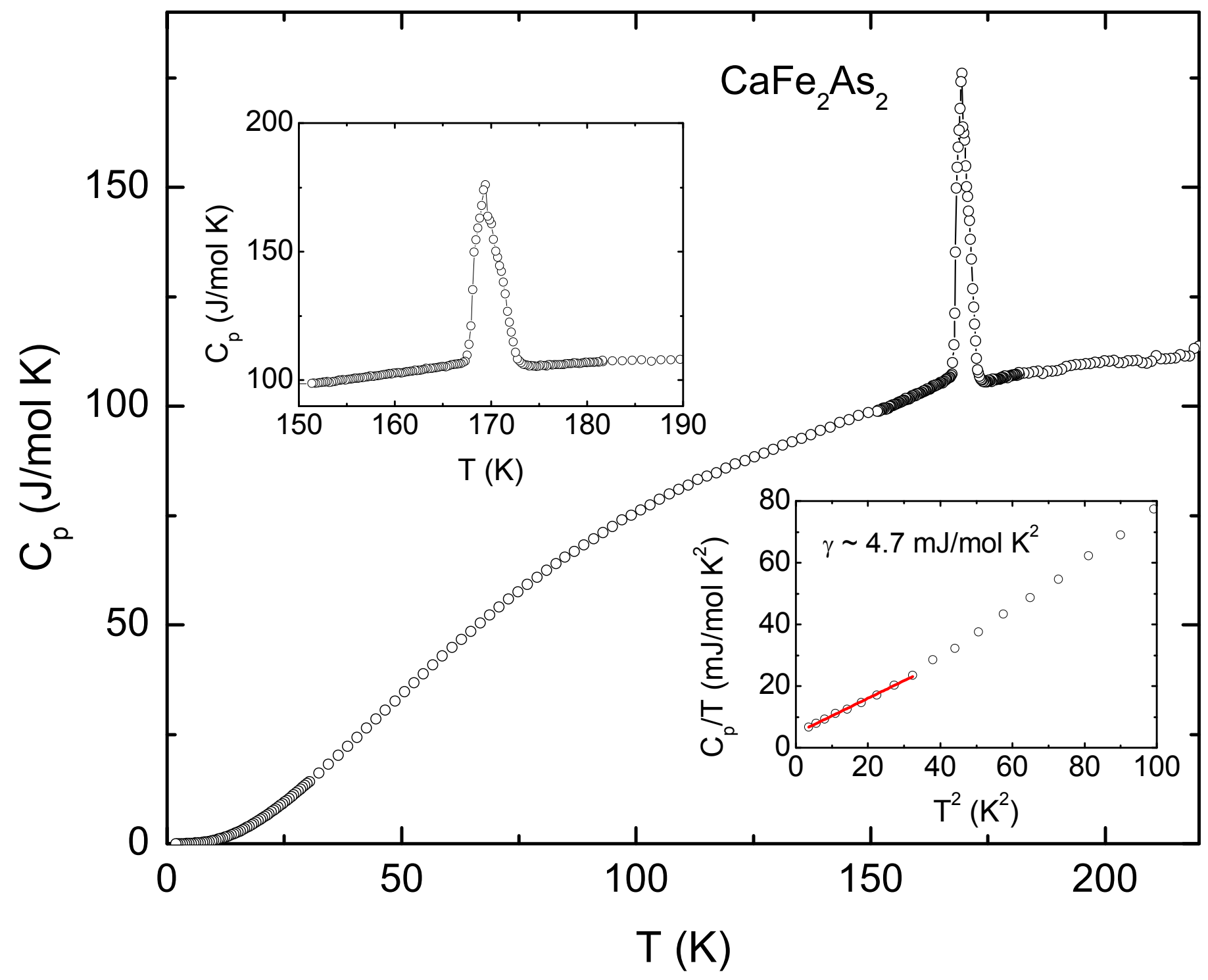

Figure 3: Temperature-dependent specific heat of $CaFe_2As_2$. Upper inset: enlargement of data in the vicinity of the 170 K phase transition. Lower inset: $C_p /T$ plotted as a function of $T^2$; line—linear fit. (After Ref. 11)

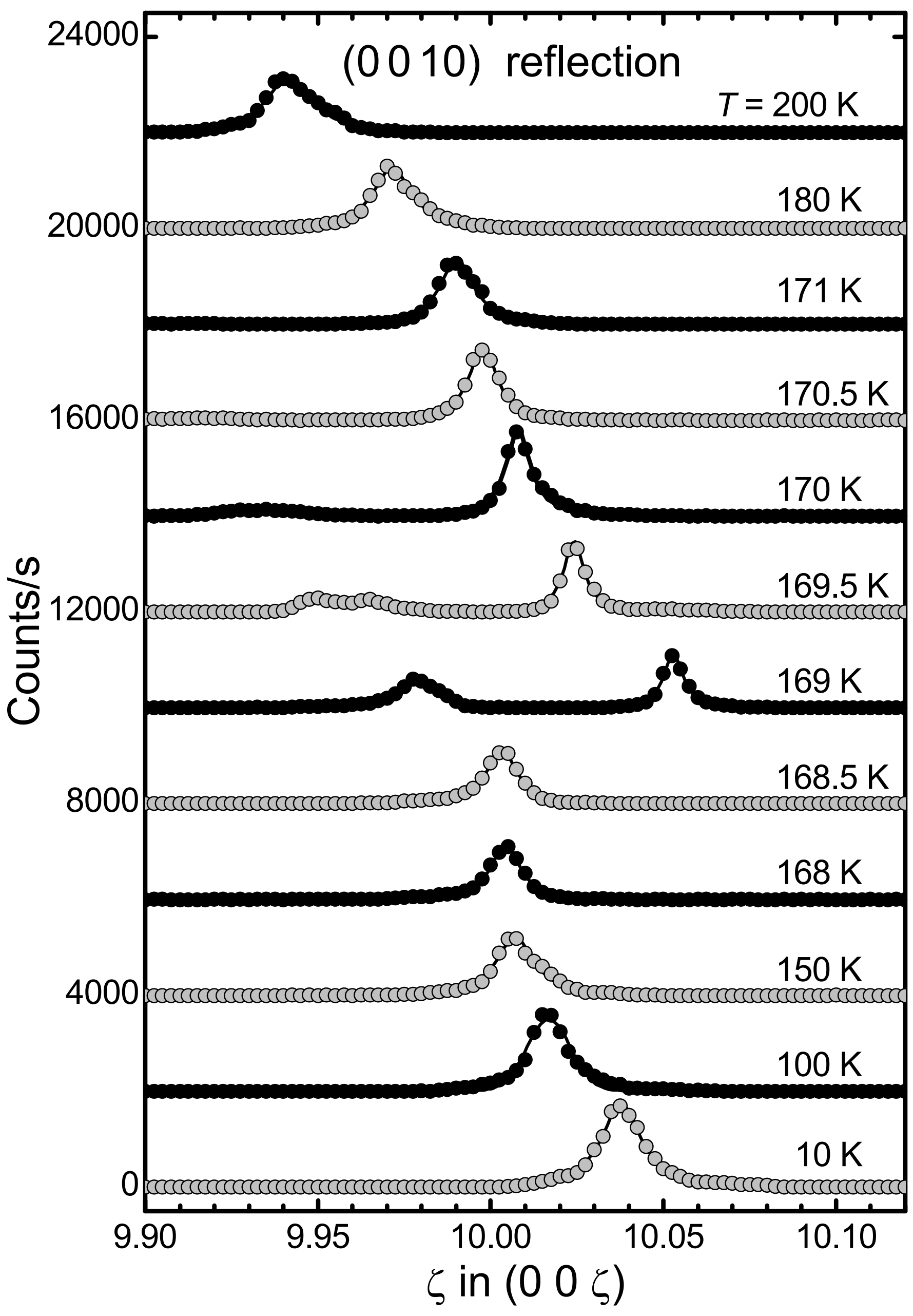

Figure 4: Longitudinal (0 0 ξ) scans through the position of the tetragonal (0 0 10) reflection for selected temperatures. The offset between every data set is 2000 counts/s. In the coexistence range, we point out that the strong change in the position of the peaks results mainly from misalignment related to the strain occurring at the phase transition in combination with the lattice-parameter changes between both phases. (After Ref. 11)

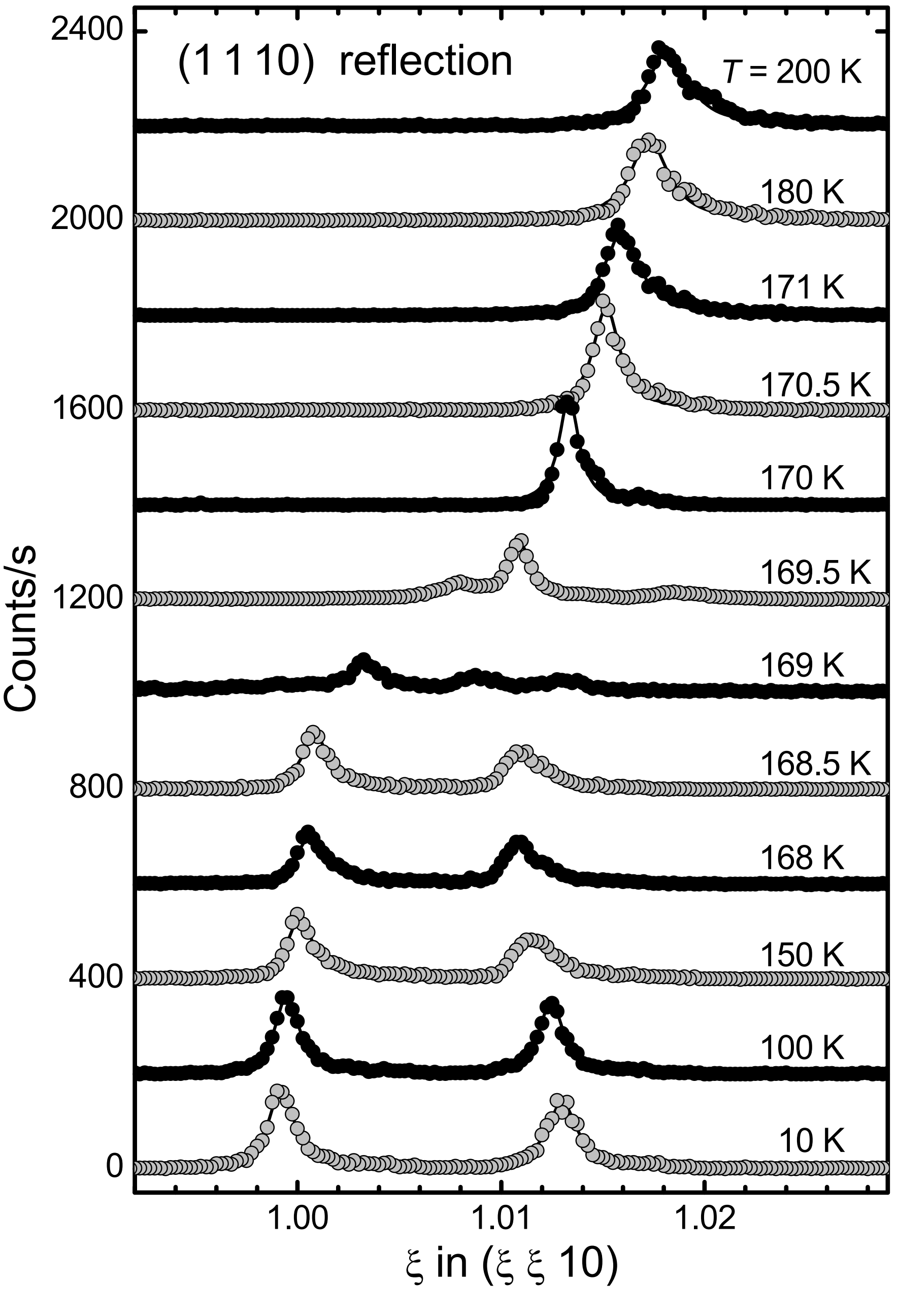

Figure 5: Transverse (ξ ξ 0) scans through the position of the tetragonal (1 1 10) reflection for selected temperatures. The offset between every data set is 200 counts/s. In the coexistence range, we point out that the strong change in the position of the peaks results mainly from misalignment related to the strain occurring at the phase transition in combination with the lattice parameter changes between both phases. (After Ref. 11)

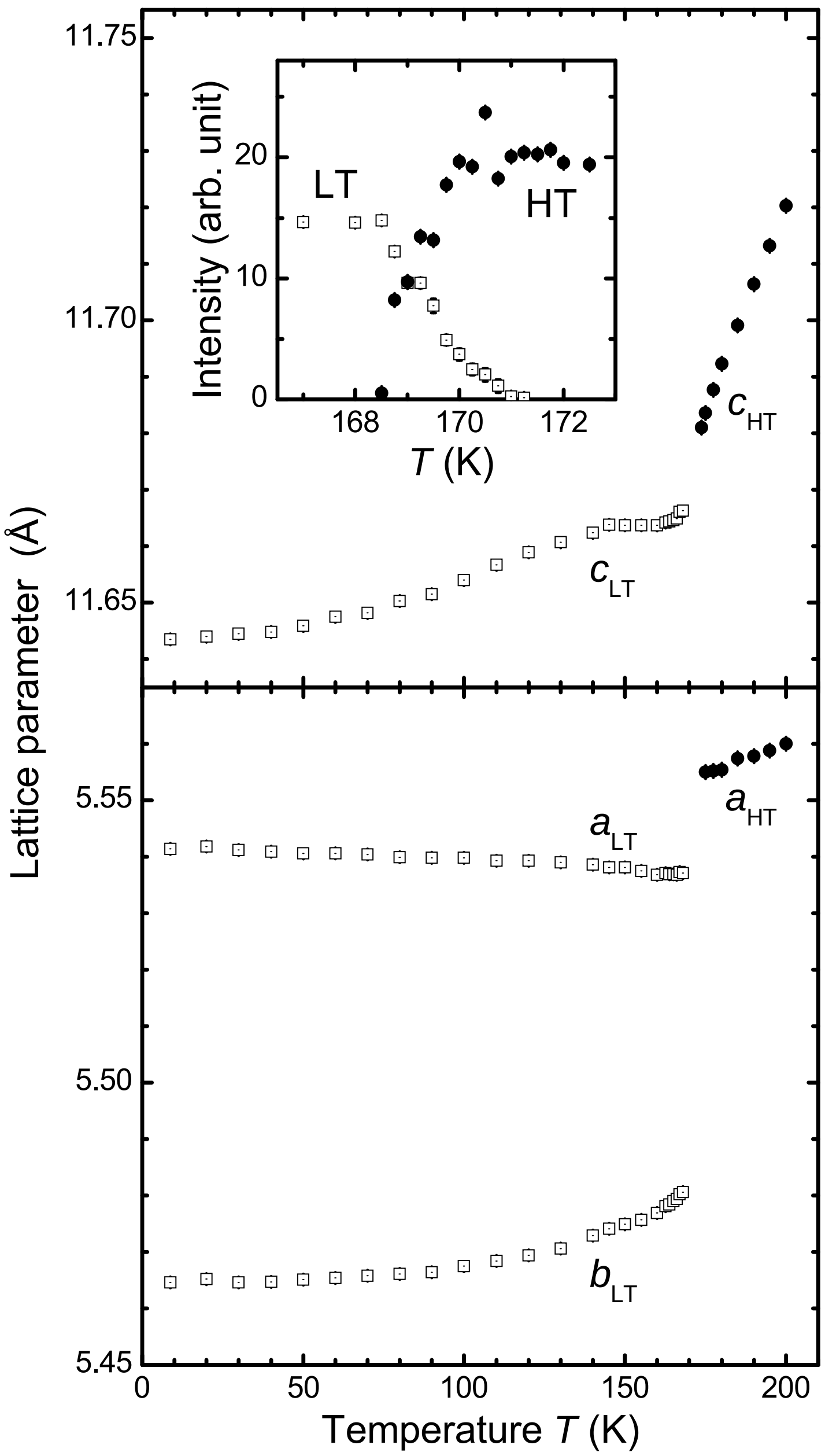

Figure 6: Lattice parameters for the tetragonal and orthorhombic phases as extracted from the data shown in Figs. 4 and 5. To allow direct comparison of both phases the lattice parameter $a$ of the tetragonal phase is given as $a_{HT}=\sqrt{2}a$. The inset shows the intensity of the reflections related to the lattice parameter $c$ shown in Fig. 4 in both phases. The error bars represent only the relative precision of the measurements. (After Ref. 11)

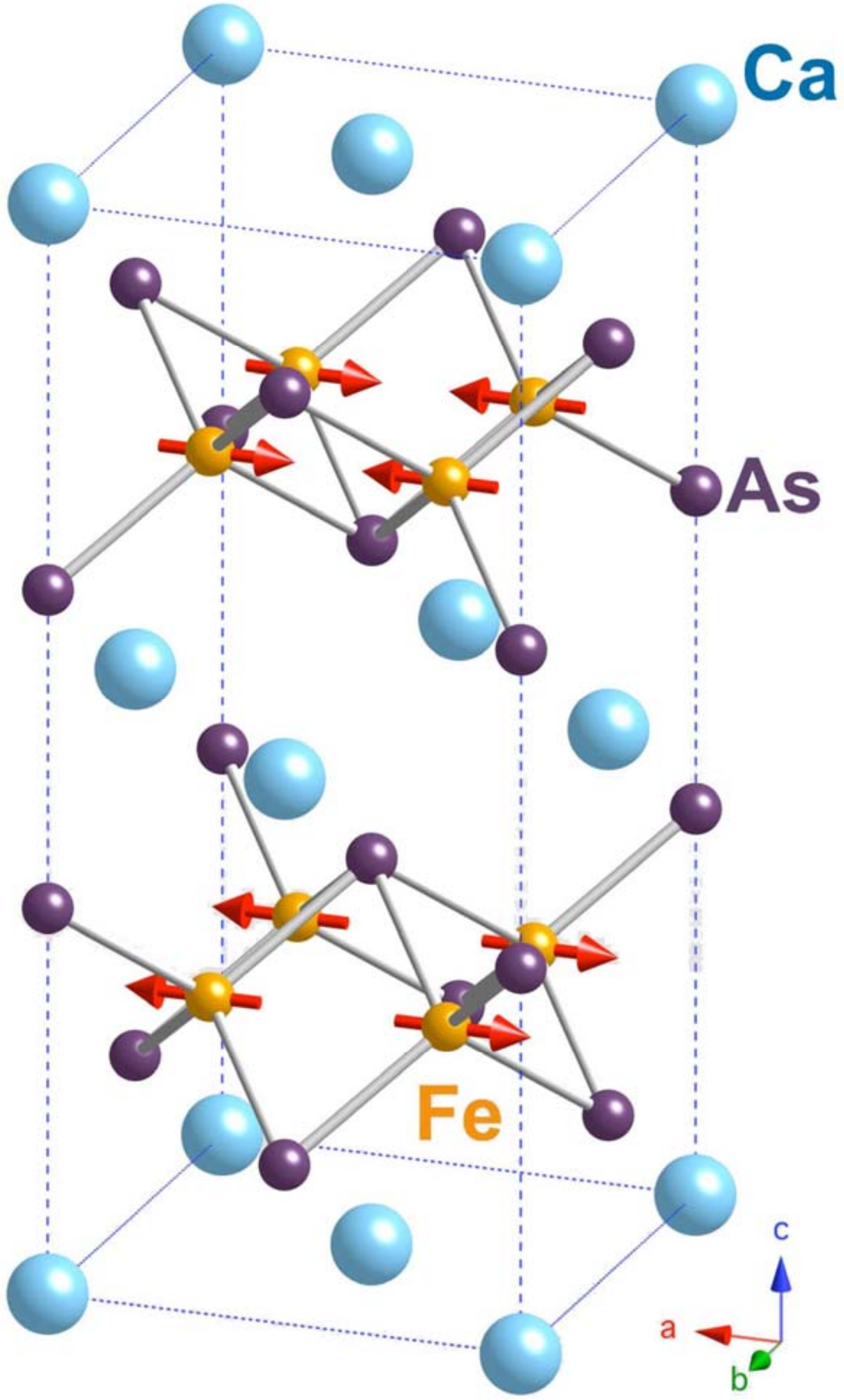

Figure 7: Illustration of the low temperature, antiferromagnetic structure of $CaFe_2As_2$. The magnetic unit cell is the same as the orthorhombic chemical unit cell. Fe moments are oriented along the orthorhombic *a* axis. (After Ref. 14)

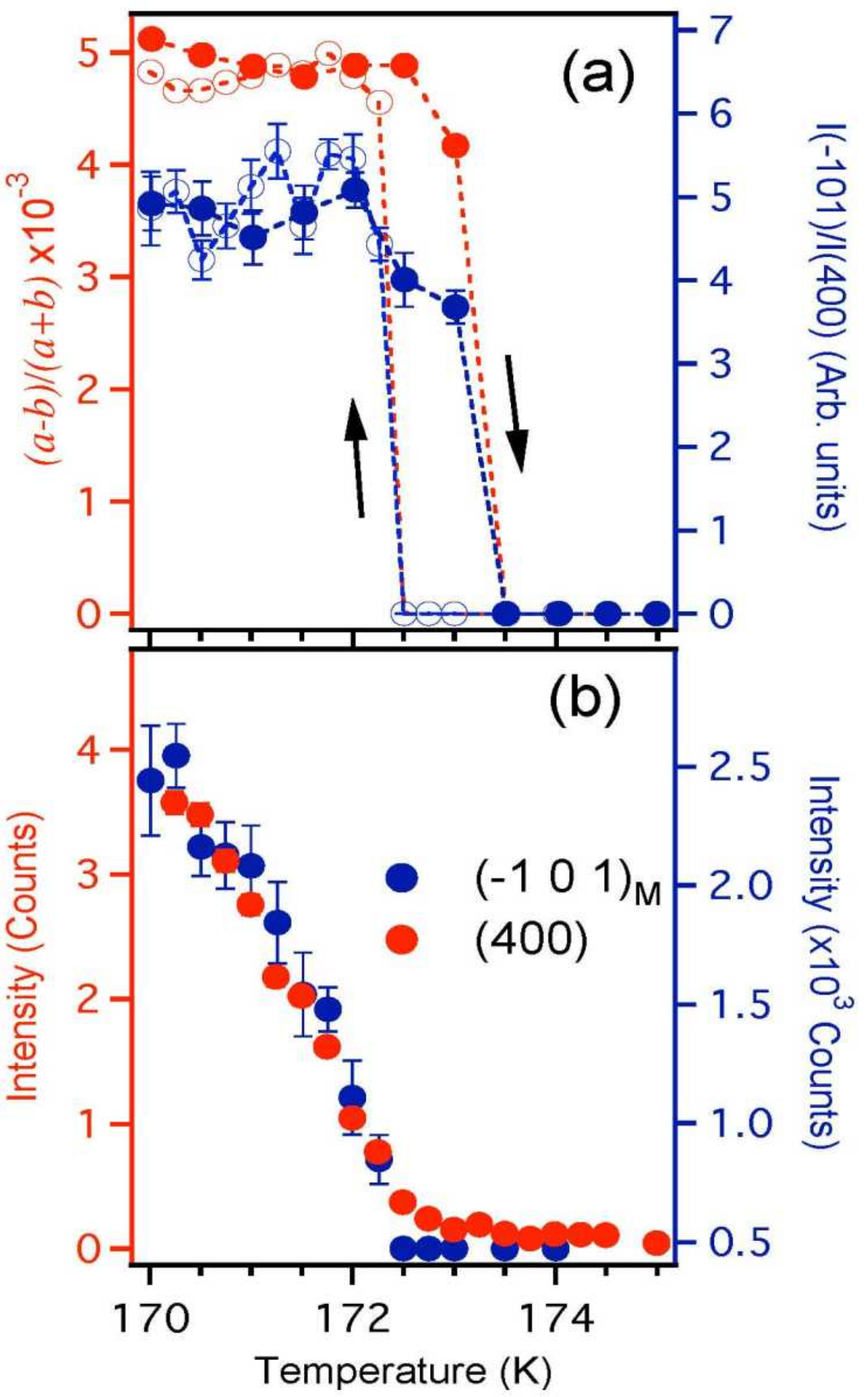

Figure 8: (a) Temperature dependence of the orthorhombic splitting (red curves) and magnetic integrated intensity (blue symbols) normalized to the orthorhombic volume fraction upon warming (filled circles) and cooling (open circles) through the transition. Below 170 K, both the orthorhombic distortion and the magnetic peak intensity are saturated. The dashed lines are guides to the eye. (b) The raw integrated intensities of the magnetic (−1 0 1) reflection (blue symbols) and the orthorhombic (4 0 0) nuclear peak. (After Ref. 14)

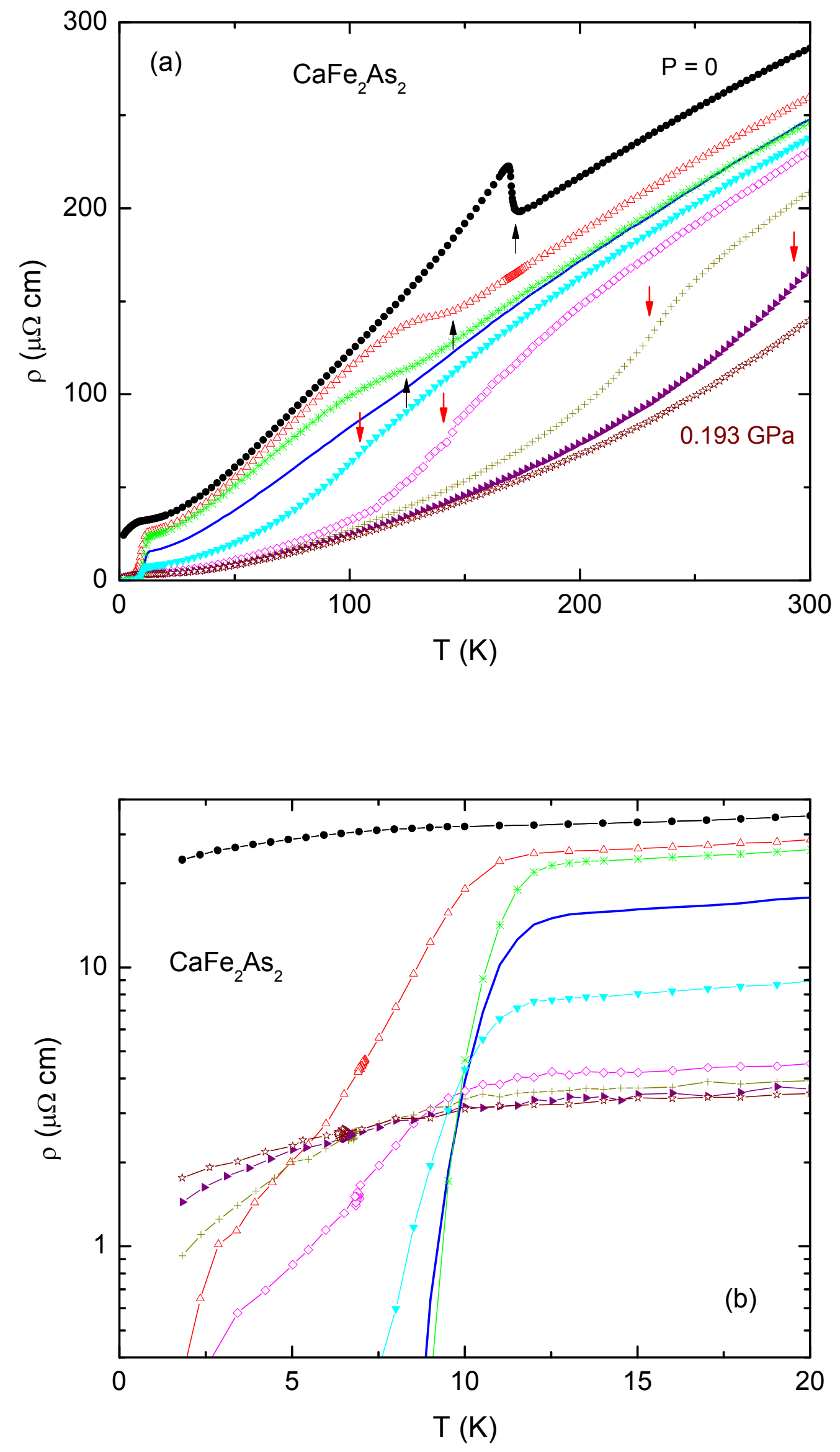

Figure 9: (a) The in-plane, electrical resistivity of $CaFe_2As_2$ as a function of temperature for *P* = 0, 0.23, 0.35, 0.51, 0.55, 0.86, 1.27, 1.68, and 1.93 GPa. The arrows indicate the location of the upper transitions temperatures. (b) Low temperature expansion of data shown in panel (a) shown on a semilog plot so as to clearly present details for all applied pressures despite a dramatic drop in the residual resistivity at higher pressures. Symbols are the same as those used in panel (a). (After Ref. 17)

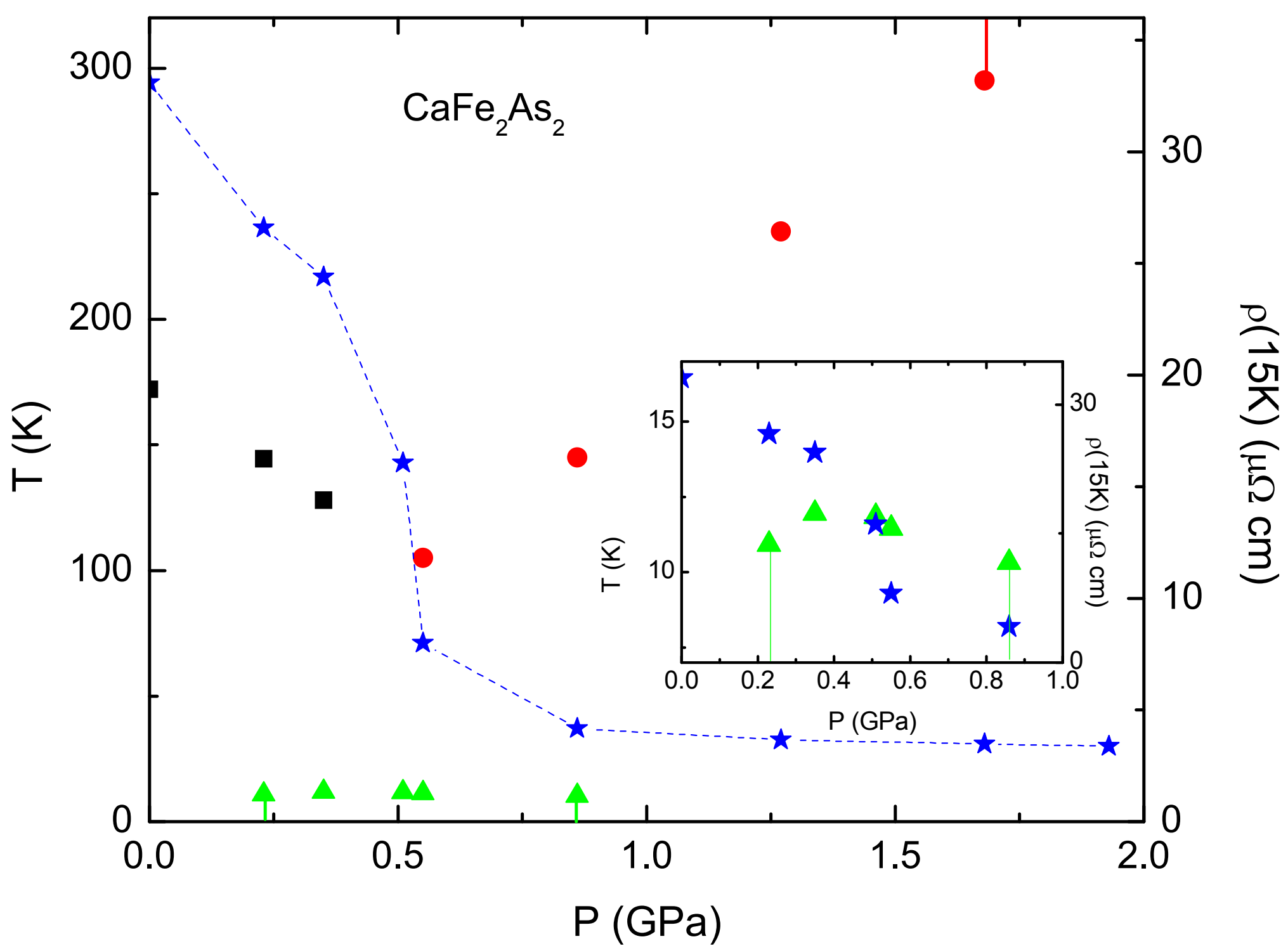

Figure 10: Pressure-temperature phase diagram of $CaFe_2As_2$ (liquid medium cell). Solid squares represent lower pressure transitions from the high temperature tetragonal phase to the low temperature orthorhombic (antiferromagnetic) phase. Solid circles represent the higher pressure phase transition that is evidenced by a marked loss of low temperature resistivity. Solid triangles represent the low temperature transition to the superconducting state. The solid stars are the 15 K resistivity values (plotted against the right-hand axis). The inset shows more clearly how the superconducting dome is centered around the sharp loss of low temperature resistivity near 0.5 GPa. (After Ref. 17)

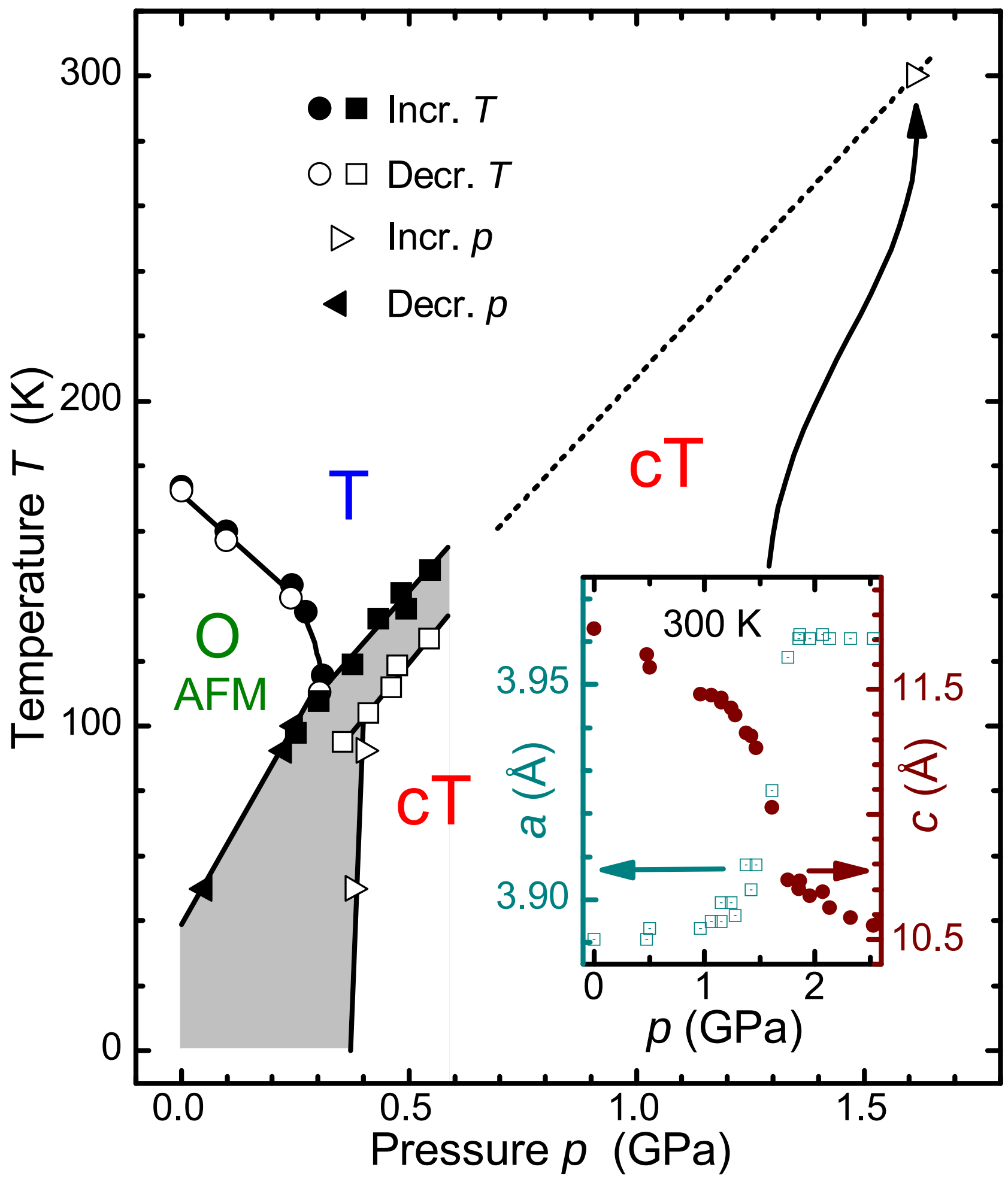

Figure 11: Pressure-temperature phase diagram of $CaFe_2As_2$ under hydrostatic pressure (He-gas cell) determined from neutron and high energy x-ray diffraction measurements. Filled and open circles (squares) denote phase boundaries determined upon heating and cooling at a set pressure for the orthorhombic – tetragonal (O-T) [collapsed tetragonal – tetragonal (cT-T)] phase transition, respectively. Filled and open triangles denote phase boundaries determined upon increasing and decreasing pressure at a fixed temperature, respectively. The shaded area denotes the hysteretic region. The inset shows the change in lattice constants at the T-cT transition at 300 K as measured by high-energy x-ray diffraction. (After Ref. 16)

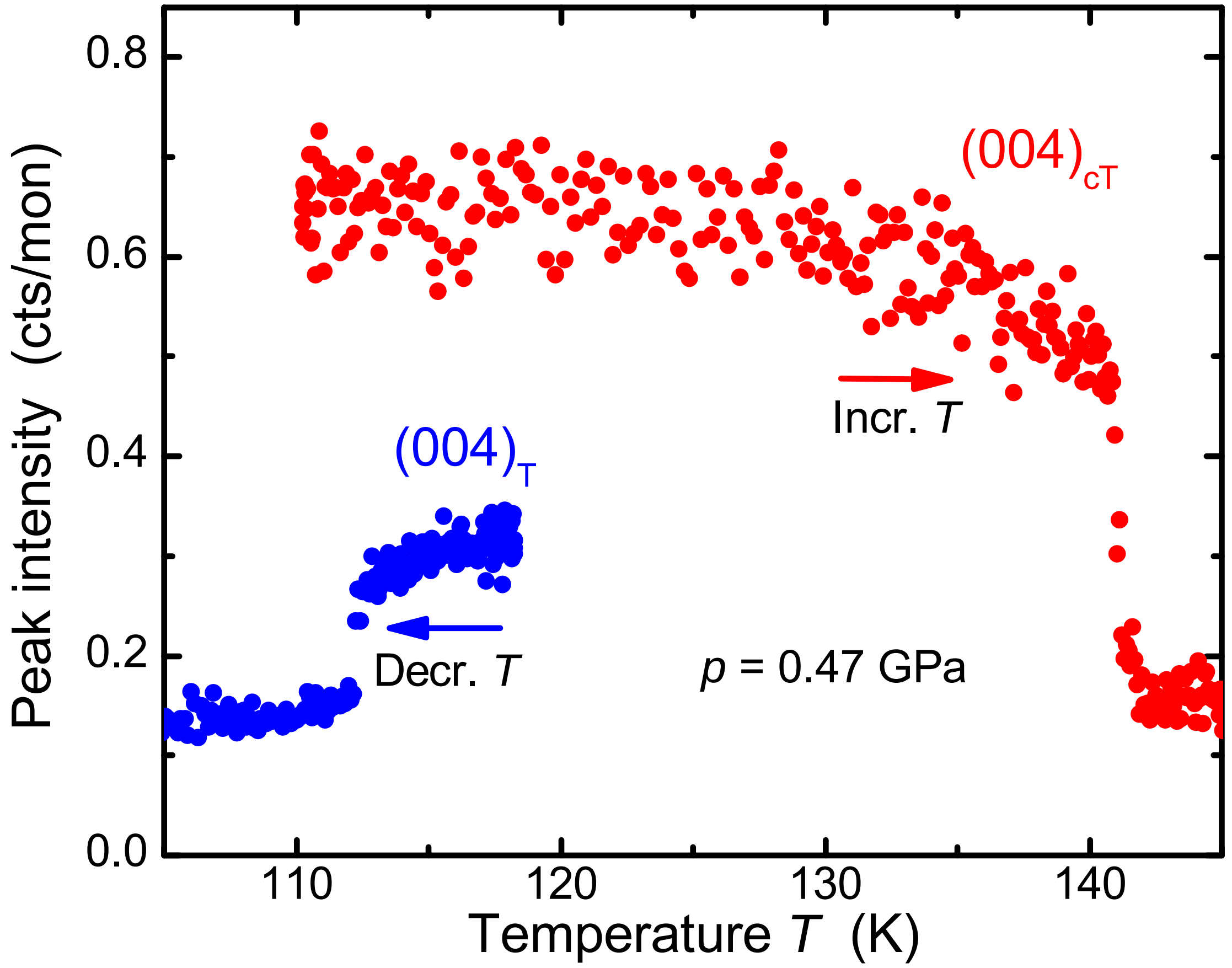

Figure 12: Temperature dependence of the peak intensities of the $(004)_T$ Bragg reflection as temperature is decreased through the T-cT transition at $P$ = 0.47 GPa, and the $(004)_{cT}$ Bragg reflection as temperature is increased through the cT-T transition at $P$ = 0.47 GPa. (After Ref. 16)

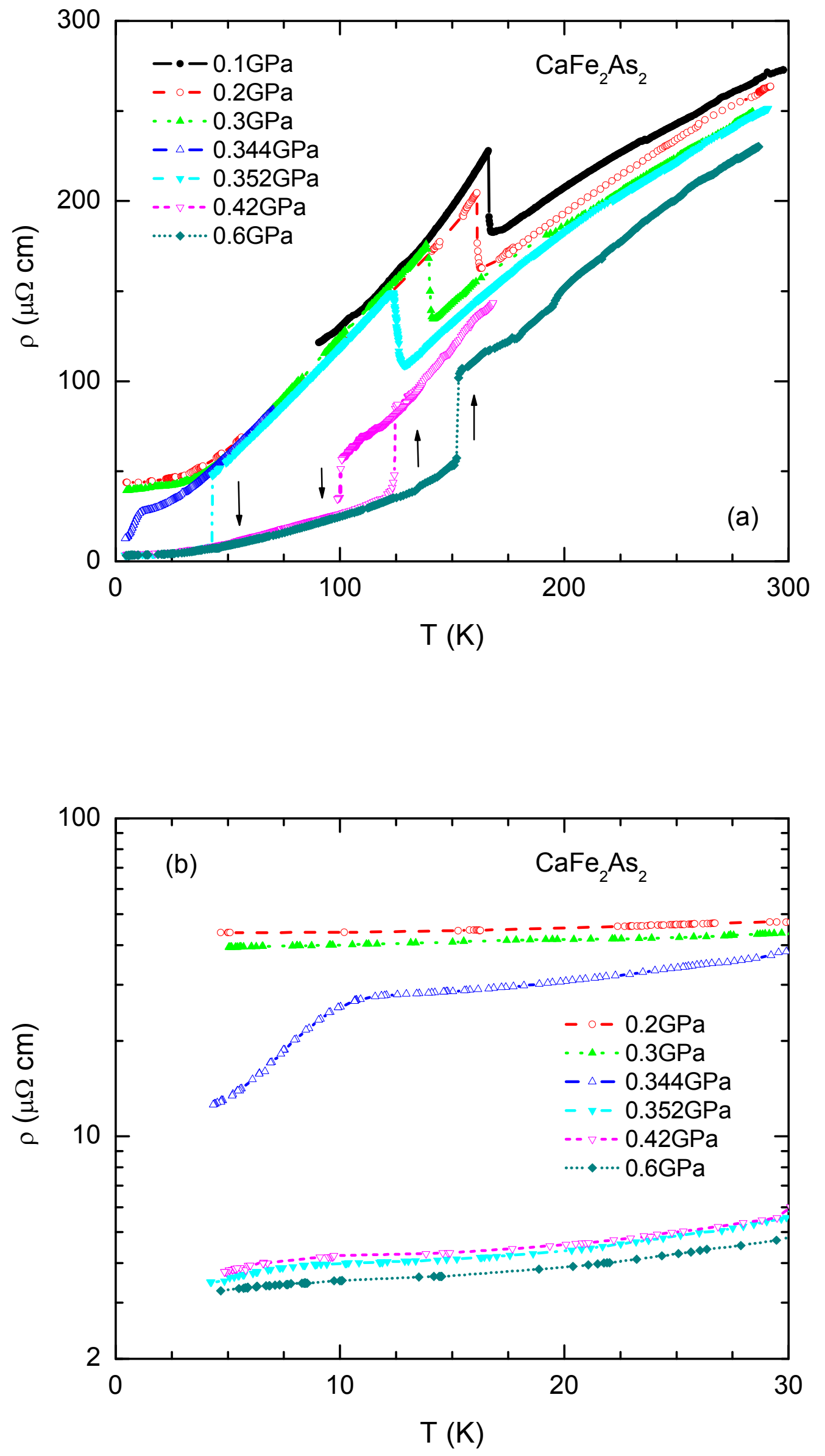

Figure 13: (a) The resistivity of a $CaFe_2As_2$ single crystal at different pressures. For pressures above 0.35 GPa, the cooling or the warming up direction is indicated by an arrow next to the plots. (b) The low-temperature resistivity of the same $CaFe_2As_2$ crystal. (After Ref. 20)

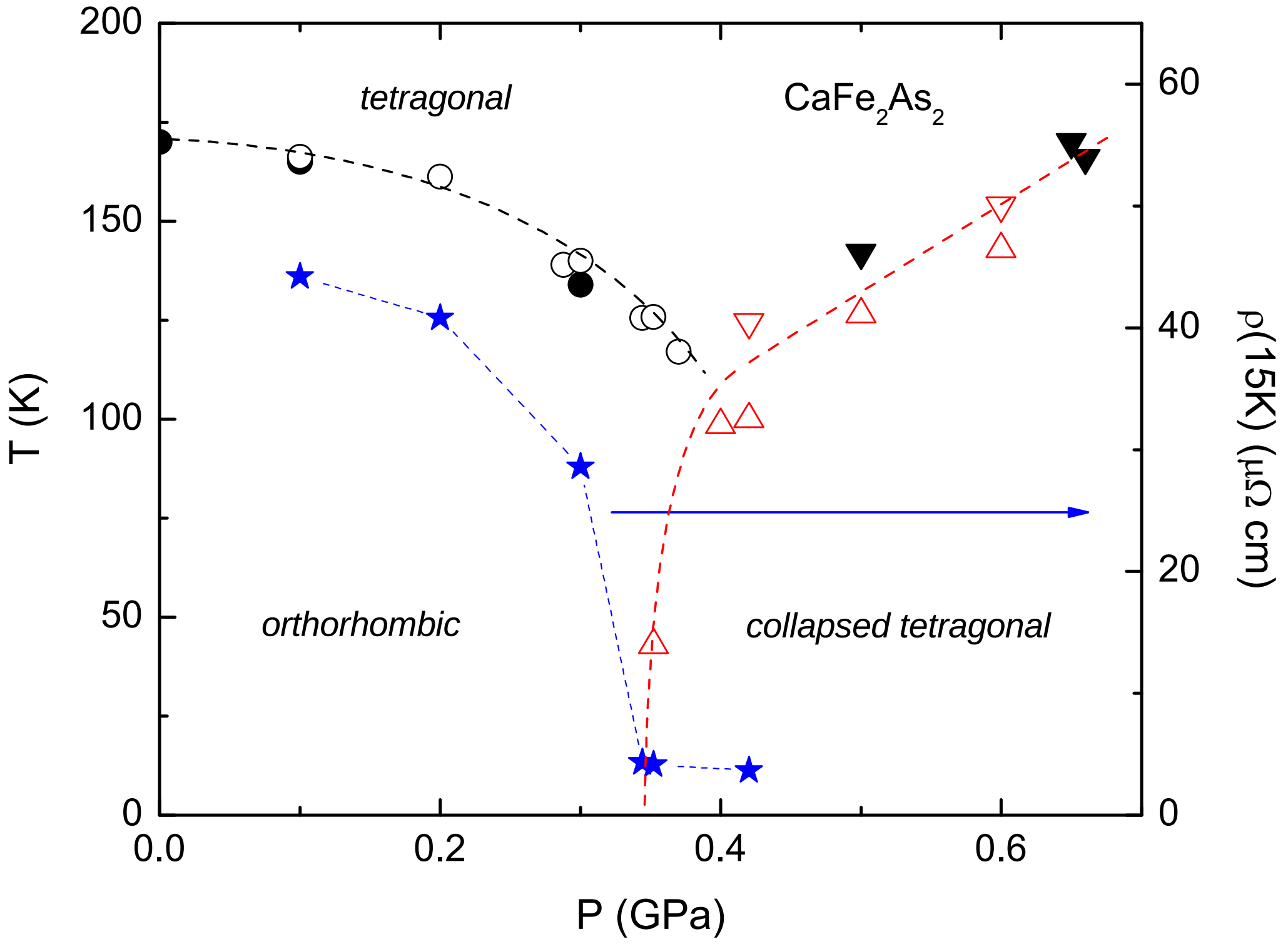

Figure 14: The *T* - *P* phase diagram of $CaFe_2As_2$ constructed from the transport and susceptibility (not shown) measurements in a He-gas pressure cell. The solid circles and the solid inverted triangles correspond to transition temperatures inferred from susceptibility data taken during warm up after ZFC. The hollow circles and the hollow inverted triangles correspond to transition temperatures inferred from transport data taken during warm up. The hollow triangles correspond to transition temperatures inferred from electrical transport data taken during cooling down. The solid stars are the 15 K resistivity values (plotted against the right-hand axis). (After Ref. 20)

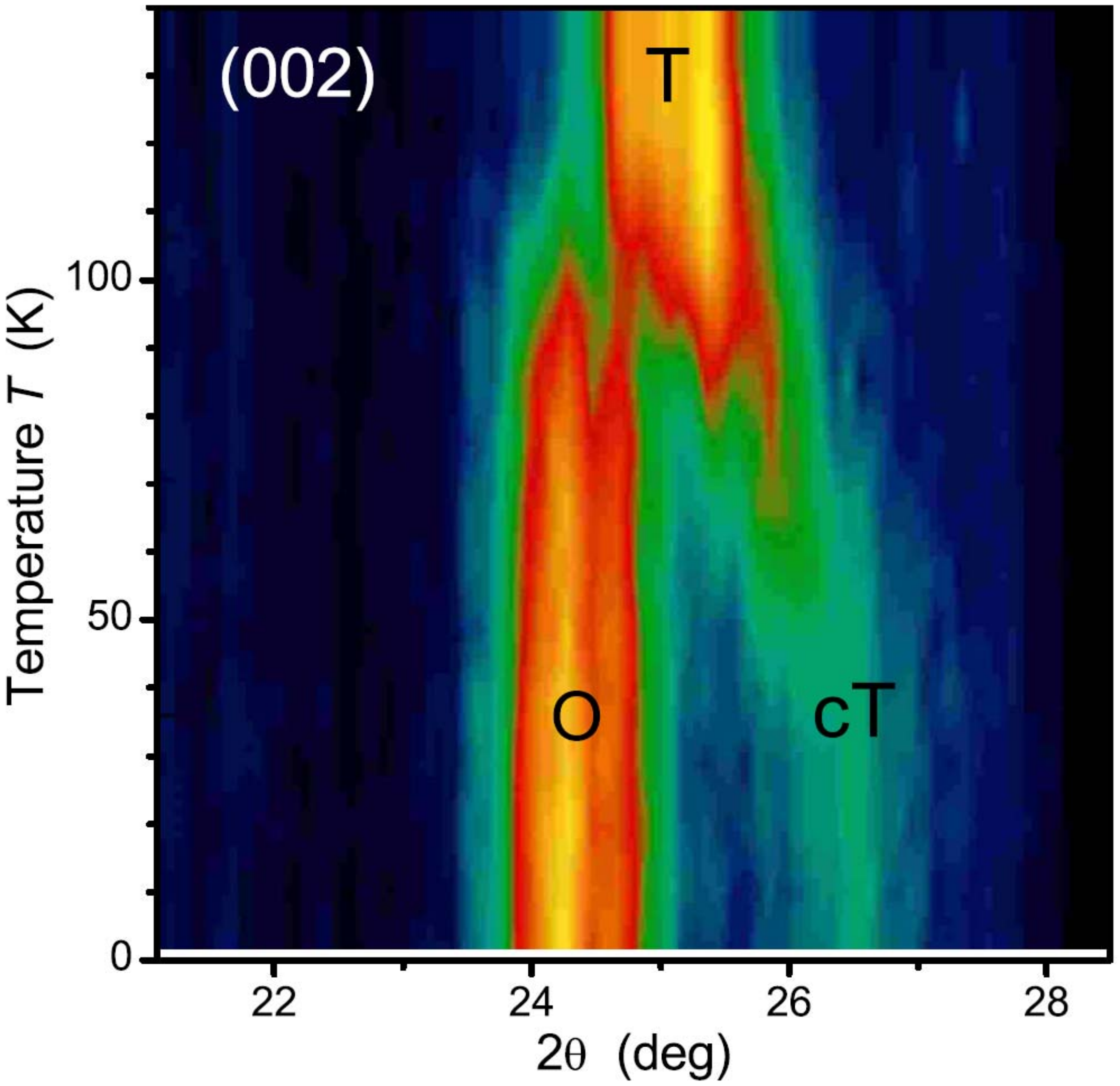

Figure 15: Measurement of the (0 0 2) nuclear reflection from $CaFe_2As_2$ with an area detector on the E4 diffractometer (Helmholtz-Zentrum) using a Be-Cu clamp-type liquid medium cell. For an initial pressure of 0.83 GPa, at room temperature, the T phase transforms to a mixture of the cT and O phase below approximately 100 K. (After Ref. 16)